\documentclass[reprint,superscriptaddress,amsmath,amssymb,aps,prb,]{revtex4-2}
\usepackage{amsmath}
\usepackage{graphicx}
\usepackage{dcolumn}
\usepackage{bm}
\usepackage{subcaption}  
\usepackage[colorlinks=true, linkcolor=blue, urlcolor=blue, citecolor=blue, pdfusetitle]{hyperref}
 
\begin{document}

\title{One-dimensional interacting Su-Schrieffer-Heeger model at quarter filling: An exact diagonalization study}

\author{Yan-Xiao Wang}
\affiliation{Key Laboratory of Quantum Theory and Applications of MoE \& School of Physical Science and Technology, Lanzhou University, Lanzhou 730000, China}
\affiliation{Lanzhou Center for Theoretical Physics, Key Laboratory of Theoretical Physics of Gansu Province, Lanzhou University, Lanzhou 730000, China}

\author{Yin Zhong}
\email{zhongy@lzu.edu.cn}
\affiliation{Key Laboratory of Quantum Theory and Applications of MoE \& School of Physical Science and Technology, Lanzhou University, Lanzhou 730000, China}
\affiliation{Lanzhou Center for Theoretical Physics, Key Laboratory of Theoretical Physics of Gansu Province, Lanzhou University, Lanzhou 730000, China}

\begin{abstract}
This study explores the ground-state phase diagram and topological properties of the spinless 1D Su-Schrieffer-Heeger (SSH) model with nearest-neighbor (NN) interactions at quarter filling. We analyze key physical quantities such as the local electron density distribution, correlation functions for bond-order-wave (BOW) and charge-density-wave (CDW) —by integrating twisted boundary conditions with the Lanczos technique and employing high-precision numerical diagonalization methods, complemented by a mean-field approximation (MFA) based on bond-order and charge-density modulation analysis. This approach enables precise identification of phase transition critical points.
Our results indicate that the system exhibits a topologically trivial band insulating (BI) phase for strong attractive interactions, with its upper boundary forming a downward-opening curve peaking at $V/t\simeq-2.3$ and extending to $V/t\simeq-2.6$. Within $-2.6 \leq V/t \leq -0.5$, a BOW phase emerges for $\left|\delta t/t\right| > 0.45$, with its boundaries converging as $\left|\delta t/t\right|$ decreases, terminating at a single point at $\left|\delta t/t\right|\simeq0.45$. In other parameter regions, a CDW phase is realized.
Through this analysis, we elucidate the topological properties of the interacting spinless SSH model at quarter filling, highlighting the competition among CDW, BOW, and BI phases. By tuning $V$ and $\delta t$, the system exhibits diverse correlated phenomena, offering new insights into one-dimensional quantum phase transitions and the interplay between topology and order.
\end{abstract}

\maketitle

\section{\label{sec:level1}INTRODUCTION}
The discovery of topological insulators (TIs) and topological superconductors (TSCs) has highlighted the profound influence of topological order on band structures. In non-interacting systems, topological invariants (e.g., the winding number and Chern number) are well-defined within Bloch state theory and directly linked to observable quantities such as edge states and quantized transport \cite{Peierls1930,Qi2011,Hasan2010}. Recent advances in universal topological classification, based on the Dirac models, demonstrate that mapping the Brillouin zone (BZ) to the target space via the winding number provides a unified framework for describing topological phases across various dimensions and symmetry classes \cite{Schnyder2008,Kitaev2009,Chiu2016,Gersdorff2021}.
In interacting systems, the absence of well-defined Bloch states complicates the definition of topological invariants. To address this, real-space methods such as the topological marker have been developed to analyze topological properties in disordered and finite-size systems \cite{Melo2023}. Additionally, entanglement spectrum degeneracy offers a powerful tool for identifying interaction-induced topological phases \cite{Legner2013,Zache2022,Rao2014}. Research demonstrates that entanglement spectrum degeneracies serve as robust indicators of topological order, undergoing distinct splitting or intersection patterns at phase transitions \cite{Pollmann2010,Pollmann2012,Turner2011,van2018}. This behavior underpins the theory of symmetry-protected topological matter, where quantum phases remain distinct under specific symmetry constraints \cite{Gu2009,Senthil2015}. Moreover, momentum-space Green’s function formalism provides an alternative approach for defining topological invariants in weakly interacting regimes, incorporating interaction effects perturbatively \cite{Gurarie2011,Essin2011}. 

In the field of one-dimensional condensed matter physics, the SSH model is a classic model for describing topological insulators, with significant theoretical and experimental value \cite{Su1979}. The model was originally employed to explain soliton excitations in conjugated polymers such as polyacetylene, and its core feature is the topologically nontrivial state induced by alternating NN hopping amplitudes, which gives rise to zero-energy modes at the boundaries \cite{Nakayama2024}. The existence of these zero-energy modes is intimately linked to the topological invariants of the system (such as the Zak phase and the winding number) \cite{WindingZakSSH2022,ProbingTopologicalWaveguide2019,DirectZakPhase2023}. The traditional SSH model primarily focuses on the non-interacting or weakly interacting regimes; however, when more complex interactions (such as Hubbard interactions or long-range Coulomb interactions) or additional topological correction terms (such as next-nearest-neighbor hoppings or periodic driving fields) are introduced \cite{Pillay2018,Li2019,Mikhail2024}, the topological phase structure of the system is significantly enriched. For instance, in the Floquet SSH model, the topological properties are modulated by the driving frequency, exhibiting dynamically induced topological edge states \cite{TopologicalPhaseTransition2023}, while in the SSH model with Hubbard interactions, the system can evolve into a Haldane phase \cite{HubbardSSH2024}.

The topological phase transition behavior of the  spinless SSH model, which belongs to the symmetry class BDI \cite{Su1979}, has been extensively studied. Notably, bosonic versions of this model, such as hard-core boson or photonic SSH systems, have also been investigated \cite{Fraxanet2022,Jin2023}.For example, in a spinless one-dimensional SSH model with NN interactions at half-filling, the topological phase diagram exhibits distinct regimes: when the interaction strength $V=0$, the sign of the dimerization parameter $\delta t$ determines whether the system exhibits a topological insulator ($\delta t<0$) or a trivial insulator ($\delta t>0$) \cite{Asboth2016}; upon introducing the NN interaction $V$, the system can undergo a topological-trivial insulator phase transition, evolving into a CDW state \cite{Zhu2022} under strong repulsive interactions or a phase separation (PS) state under strong attractive interactions \cite{Melo2023}.

\begin{figure}
     \centering
     \includegraphics[width=0.5\textwidth]{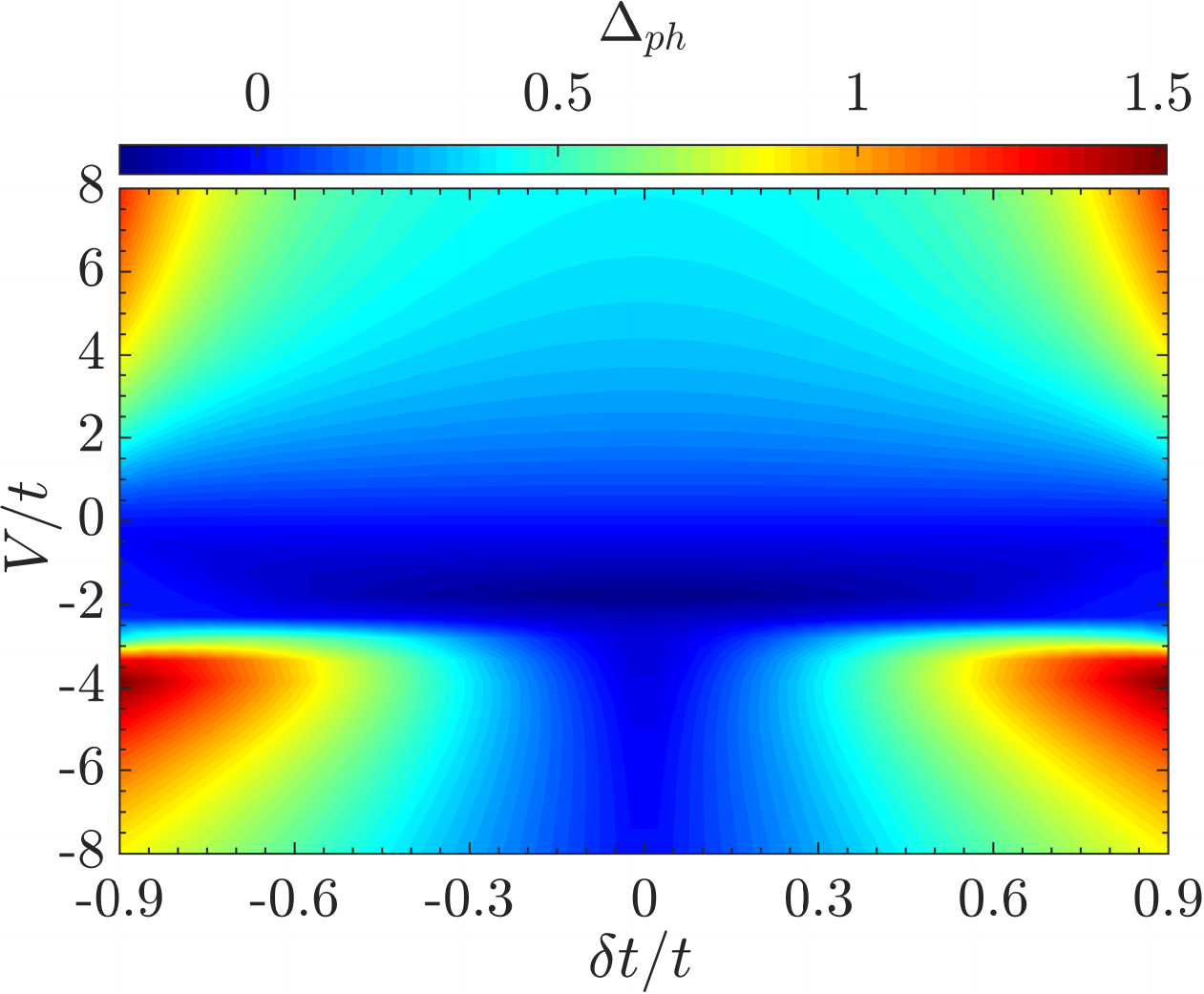}
     \caption{The distribution of the particle-hole energy gap $\Delta_{ph}$ in the parameter space $(\delta t, V)$ reveals the metal-insulator transition of the system. At $V = 0$, $\Delta_{ph}$ vanishes, indicating a gapless metallic phase. As the interaction strength $V$ increases, $\Delta_{ph}$ becomes finite, signaling the onset of an insulating phase characterized by a finite excitation gap. The occurrence of negative values of $\Delta_{ph}$ in the figure originates from finite-size effects. And the expression used to calculate $\Delta_{ph}$ here is derived from Eq.~\ref{eq3}.}
     \label{fig1}
   \end{figure}

In contrast, the phase diagram of the SSH model at quarter filling remains relatively unexplored. At this filling, the reduced electron density leads to a substantial shift in the Fermi surface compared to the half-filled case. Moreover, the alternating hopping amplitudes $t\pm\delta t$, characteristic of the SSH model, no longer fully open a gap in the single-particle spectrum, resulting in gapless excitations and quasi-metallic behavior consistent with a Luttinger liquid in one-dimensional systems. However, the introduction of NN interactions $V$ significantly enhances correlation effects and charge fluctuations due to the low carrier density. As a result, the system may undergo a correlation-driven metal-to-insulator transition, as illustrated in Fig.~\ref{fig1}. Compared with the half-filled case, the altered Fermi surface at quarter filling reshapes the relevant energy scales and enhances the competition among various interaction-driven orders \cite{Kohn1959}. As a result, the dominant mechanisms underlying topological phase transitions shift from being solely determined by the winding number to being strongly influenced by the interplay between different symmetry-breaking tendencies.

Motivated by these distinctions, we investigate the ground-state phase diagram of the one-dimensional spinless SSH model at quarter filling. Specifically, we address whether the reduction in filling suppresses topological properties and induces new ordered phases, and whether the critical behaviors of phase transitions differ from those observed at half filling. Given the breakdown of Bloch states under strong interactions, this study adopts a multi-order parameter approach—computing CDW and BOW order parameters and their respective correlation functions \cite{Shao2019,Tanjaroon2023,Weber2024}—to identify phase boundaries. Compared to topological markers and many-body Chern number diagnostics \cite{Varney2011,Varney2010,Kourtis2014,Shao2021}, this strategy is more suitable for strongly interacting systems and provides a clear identification of correlation-driven ordered phases.

The model and the definitions of the order parameters are introduced in Sec.~\ref{sec:level2}. Section~\ref{sec:level3} presents the numerical results based on exact diagonalization (ED), while Sec.~\ref{sec:level4} compares these results with those obtained from the MFA. Finally, the conclusions of this study and potential directions for future research are summarized in Sec.~\ref{sec:level5}.
\section{\label{sec:level2}MODEL AND METHOD}
\begin{figure}[htbp]
    \centering
    \begin{subfigure}[b]{0.42\textwidth}  
        \centering
        \includegraphics[width=\linewidth]{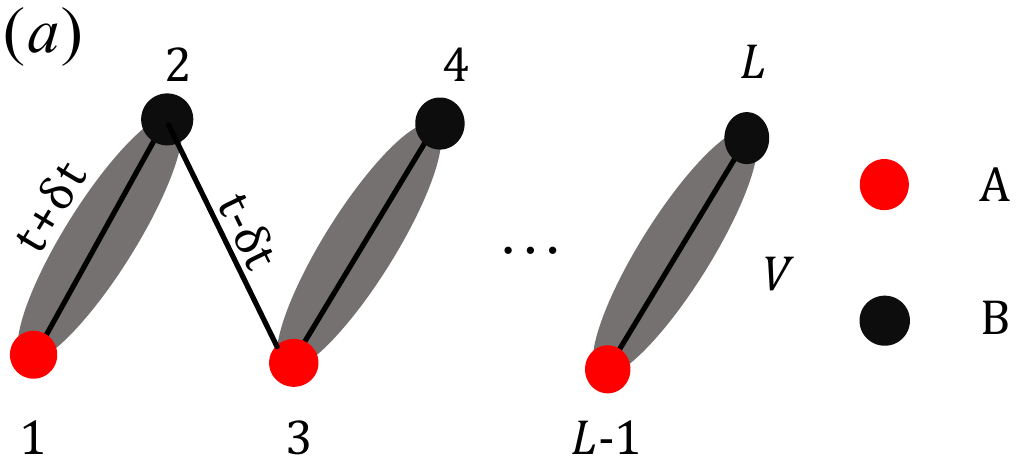} 
        \label{fig:subfig1}
    \end{subfigure}
    \hspace{0.05\textwidth}  
    \begin{subfigure}[b]{0.45\textwidth}
        \centering
        \includegraphics[width=\linewidth]{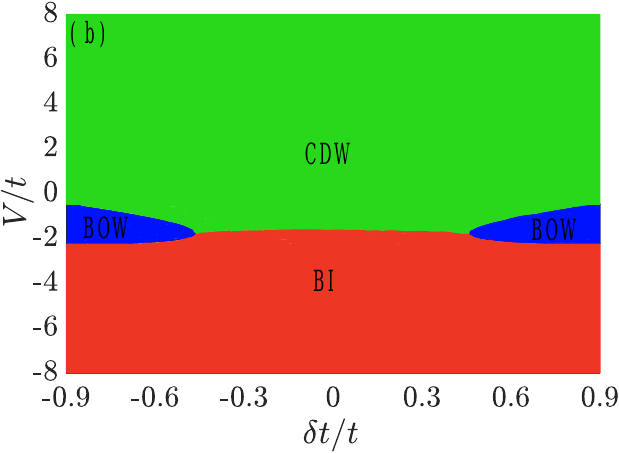}
        \label{fig:subfig2}
    \end{subfigure}
    \caption{
        Schematic representation of the Hamiltonian (a); and (b) presented is the ground-state phase diagram of this one-dimensional SSH model at quarter filling.}
    \label{fig2}
\end{figure}

In this paper, we investigate the ground state properties of a one-dimensional spinless Su-Schrieffer-Heeger  model \cite{Su1980,Su1979} with NN interactions \cite{Fradkin1983,Sirker2014,Yu2016,Yahyavi2018,Zegarra2019}.Serving as a paradigmatic system in topological physics, the Hamiltonian of the SSH model can be expressed as
    \begin{equation}\label{eq1}
       \hat{H}=\sum_{\left\langle i,j\right\rangle}\left(-t+\left(-1\right)^{i}\delta t\right)\left(\hat{c}_j^\dagger\hat{c}_i+h.c.\right)+V\sum_{\left\langle i,j\right\rangle}\hat{n}_i\hat{n}_j.
    \end{equation}
Here, $\hat{c}_i^\dagger \left( \hat{c}_i \right)$ denotes the creation (annihilation) operator for fermions, where $i$ and $j$ denote sites of the lattice with length $L$. The summation $\left\langle i,j \right\rangle$ is restricted to NN sites. The parameter $t$ represents the hopping integral, whose magnitude is modulated by the dimerization parameter $\delta t$, and $V$ is the nearest-neighbor interaction. The particle number operator at the $i$-th site is defined as $\hat{n}_i = \hat{c}_i^\dagger \hat{c}_i$ and the sublattices within a unit cell are labeled by the subscripts A and B, as illustrated in Fig.~\ref{fig2}(a). In subsequent discussions, we normalize the lattice constant $a$ to unity ($a \equiv 1$) and adopt $t$ as the unit of energy, i.e., all energy scales are measured in units of $t$.

In order to investigate the ground-state properties at quarter filling (i.e., $N=L/4$), the ED method \cite{Lin1993} is employed to numerically solve the Eq.~ \ref{eq1} for a system size of $L=16$. To mitigate finite-size effects, we introduce twisted-averaged boundary conditions (TABC) \cite{Poilblanc1991,Gros1992,Lin2001,Zawadzki2017}, whereby a phase correction is applied to the hopping integral when fermions hop between the first and last sites: $\phi:t_{ij}\rightarrow t_{ij}\exp i\phi_{l}$, where $\phi_{l}=\left[0,2\pi\right)$. Each phase $\phi_{l}$ corresponds to a different set of $k$-points, effectively increasing the number of accessible momentum points within the first Brillouin zone for a given lattice configuration, thereby suppressing finite-size effects. Within this framework, the expectation value of an operator $\hat{A}$ is given by \cite{Koretsune2007}
    \begin{equation}\label{eq2}
       \left\langle\hat{A}\right\rangle=\frac{1}{N_{\phi}}\sum_{l=1}^{N_{\phi}}\left\langle\hat{A}\right\rangle_{\phi_{l}},
    \end{equation}
in this study, we set the number of phase sampling points to $N_{\phi}=20$.

To characterize the metallic or insulating nature of the system, we calculate the particle-hole excitation gap, defined as
    \begin{equation}\label{eq3}
       \Delta_{ph}\equiv E_{N+1}+E_{N-1}-2E_{N},
    \end{equation}
where $N=L/4$, and $E_{N+1}$, $E_{N-1}$, and $E_{N}$ represent the ground state energies of systems with $N+1$, $N-1$, and $N$ particles, respectively. 
We also compute the single-particle addition gap, given by \cite{Mikhail2024}
    \begin{equation}\label{eq4}
       \Delta\equiv E_{N+1}-E_{N},
    \end{equation}
which quantifies the energy cost required to add a particle to the system. While $\Delta$ provides insight into single-particle excitations, it does not by itself determine whether the system is metallic or insulating. Analogous to the band gap in non-interacting systems, the relevant indicator for an insulating state is the particle-hole gap $\Delta_{ph}$, which can also be expressed in terms of the addition energy differences
   \begin{align}\label{eq5}
         \Delta_N &= \Delta(N) - \Delta(N-1) \notag\\
         &= E_{N+1} - E_N - (E_N - E_{N-1}) \notag\\
         &\equiv \Delta_{ph}.
   \end{align}
   
Characterizing the charge-order modulations driven by interactions, we numerically compute the CDW order parameter and its associated structure factor. The CDW order parameter is defined as
    \begin{equation}\label{eq6}
       O_{cdw}=\frac{1}{L}\sum_{i}e^{iqR_i}\left\langle\hat{n}_i\right\rangle, q=\frac{\pi}{2}.
    \end{equation}
Here, $R_i$ denotes the position of the i-th site. This quantity reflects the amplitude of the staggered charge density modulation, and a non-zero value indicates the breaking of translational symmetry. The corresponding structure factor is given by \cite{Melo2023}
    \begin{equation}\label{eq7}
       S_{cdw}=\frac{1}{L}\sum_{i,j}e^{iq(R_i-R_j)}\left\langle\hat{n}_i\hat{n}_j\right\rangle, q=\frac{\pi}{2}.
    \end{equation}
We also compute the real-space density-density correlation function to analyze the spatial decay behavior of the CDW order
    \begin{equation}\label{eq8}
       N_{ij}=\left\langle\hat{n}_i\hat{n}_j\right\rangle-\left\langle\hat{n}_i\right\rangle\left\langle\hat{n}_j\right\rangle.
    \end{equation}
    
In turn, to further characterize the fluctuations of the CDW in momentum space, we also compute the CDW correlation function via Fourier transform
    \begin{equation}\label{eq9}
       \chi_{cdw}\left(q\right)=\frac{1}{L}\sum_{i,j}e^{iq\left(R_i-R_j\right)}\left\langle\hat{n}_i\hat{n}_j\right\rangle,
    \end{equation}
at quarter filling, the CDW exhibits a modulation with a periodicity of four sites, corresponding to the characteristic wave vectors $q=\pm\frac{\pi}{2}$. In the presence of long-range order, the CDW susceptibility $\chi_{cdw}(q)$ displays pronounced peaks at these wave vectors, signaling the establishment of a well-defined CDW phase.

On the other hand, due to the alternating hopping strengths induced by the dimerization parameter $\delta t$, it is physically meaningful to partition the one-dimensional chain into sublattices A (odd sites) and B (even sites). This partitioning can potentially give rise to bond-order wave . To characterize such order, we construct the BOW order parameter as
    \begin{equation}\label{eq10}
       O_{bow}=\frac{1}{L}\sum_{i}e^{iqR_i}\left(\left\langle\hat{c}_i^\dagger\hat{c}_{i+1}\right\rangle-\left\langle\hat{c}_{i-1}^\dagger\hat{c}_i\right\rangle\right), q=\frac{\pi}{2}.
    \end{equation}
A non-zero value of $O_{bow}$ indicates a significant difference between the bond strengths on odd and even bonds, and it saturates to a maximum value in the ideal dimerization limit. Correspondingly, we introduce the BOW correlation function in momentum space as \cite{Shao2019,Tanjaroon2023,Weber2024,Pinaki2002}
    \begin{equation}\label{eq11}
       \chi_{bow}\left(q\right)=\frac{1}{L}\sum_{i,j}e^{iq\left(R_i-R_j\right)}\left\langle\hat{B}_i\hat{B}_j\right\rangle,
    \end{equation}
here $\hat{B}_i=\hat{c}_i^\dagger\hat{c}_{i+1}+h.c.$ is the bond strength operator, where $h.c.$ denotes the Hermitian conjugate term. Upon entering the BOW phase, the BOW susceptibility $\chi_{bow}(q)$ exhibits a divergence at $q=\pi$, reflecting the establishment of long-range or quasi-long-range bond-order correlations with a periodicity of 2 $\left(\lambda=2\right)$ .

Finally, we also characterize the quantum entanglement properties of the system by computing the von Neumann entropy, defined as \cite{Kallin2009,Chung2011}
    \begin{equation}\label{eq13}
       S_{\nu N}\left(\delta t,V\right)=\operatorname{Tr}\rho_{A}\ln\rho_{A},
    \end{equation}
to quantify the entanglement between subsystems, we select subsystem A as all the sites belonging to sublattice A. The reduced density matrix $\rho_A$ is obtained by tracing out the degrees of freedom of sublattice B from the total density matrix.
\section{\label{sec:level3}RESULTS}
After defining the primary physical quantities, we systematically analyzed the various phases of model \ref{eq1} within the parameter space $(\delta t, V)$  in Fig.~\ref{fig2}(b). Numerical results indicate that, aside from the occurrence of a BI phase at sufficiently negative interaction strengths, a BOW phase emerges in the region $-2.6 \leq V/t \leq -0.5$ with $\left|\delta t/t\right| > 0.45$. Under larger positive $V/t$, pronounced CDW characteristics are observed. The transition from BI to ordered phases (CDW or BOW) reflects a shift from a state preserving translational symmetry to one where this symmetry is broken. 
\subsection{\label{subsec:level1}Energy gap}
\begin{figure*}
      \centering
      \includegraphics[width=0.9\textwidth]{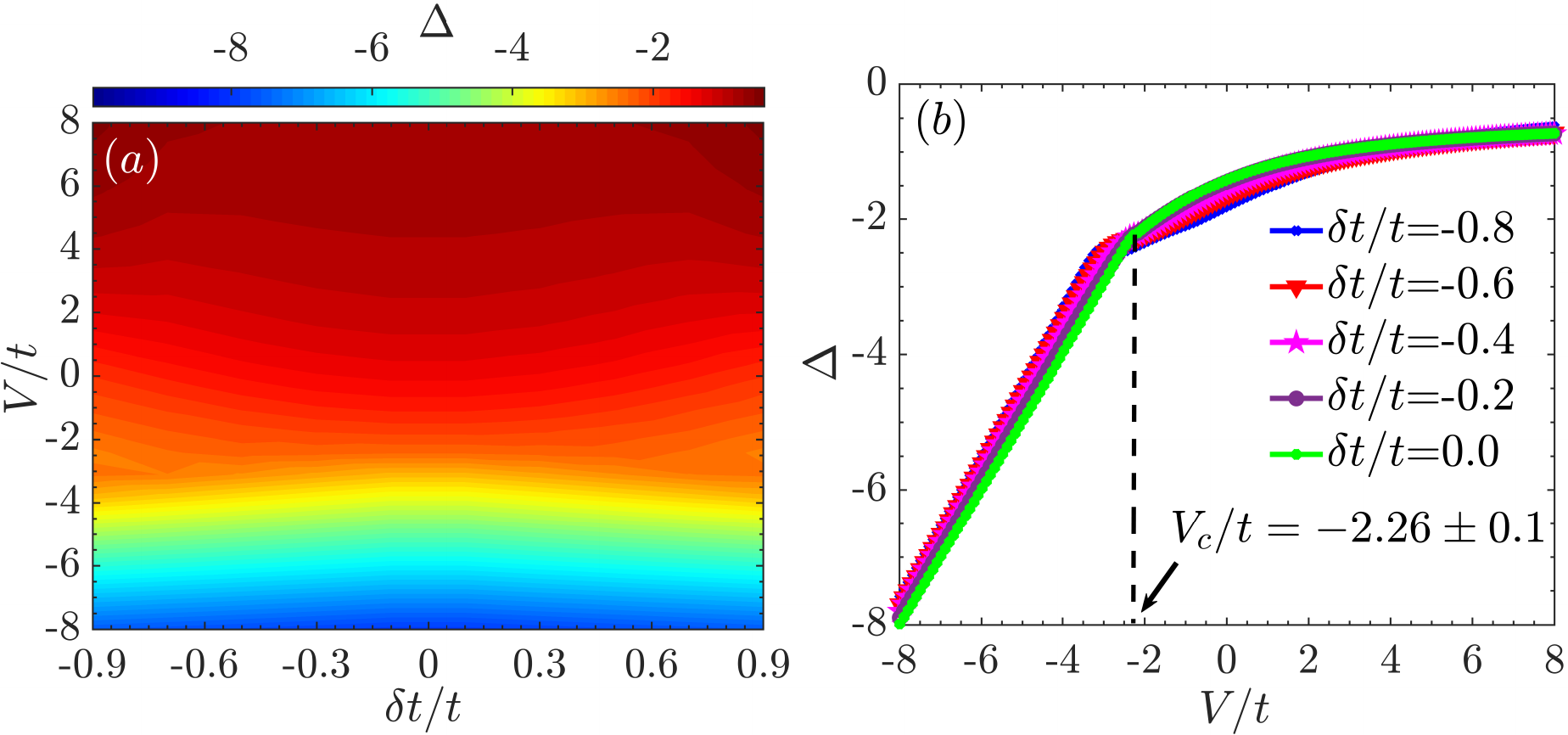}
      \caption{Panel $\left(a\right)$ displays the distribution of the energy gap $\Delta$ in the parameter space of $\left(\delta t,V\right)$.Panel $\left(b\right)$ shows the variation of  $\Delta$ as a function of $V/t$ for fixed values of $\delta t/t$. The data were computed at quarter filling with a lattice size of $L=16$.}
      \label{fig3}
    \end{figure*}  
    
We first compute the distribution of the energy gap $\Delta$ in the parameter space $\left(\delta t,V\right)$ based on Eq.~\eqref{eq4}, as shown in Fig.~\ref{fig3}(a). And (b) displays the variation of the $\Delta$ as a function of $V/t$ for fixed values of $\delta t/t$. A clear discontinuity in the slope of $\Delta$ is observed when $V/t$ approaches -2.26, with $\delta t/t=0$ at this point. Furthermore, as the absolute value of  $\delta t/t$ increases, the critical value $V_{c}$ at which the slope discontinuity occurs gradually decreases (its absolute value increases). Additionally, it is important to note that most physical quantities are symmetric about $\delta t/t=0$, and therefore, in the subsequent analysis, we focus only on the case of $\delta t/t<0$.
   \begin{figure}
     \centering
     \includegraphics[width=0.5\textwidth]{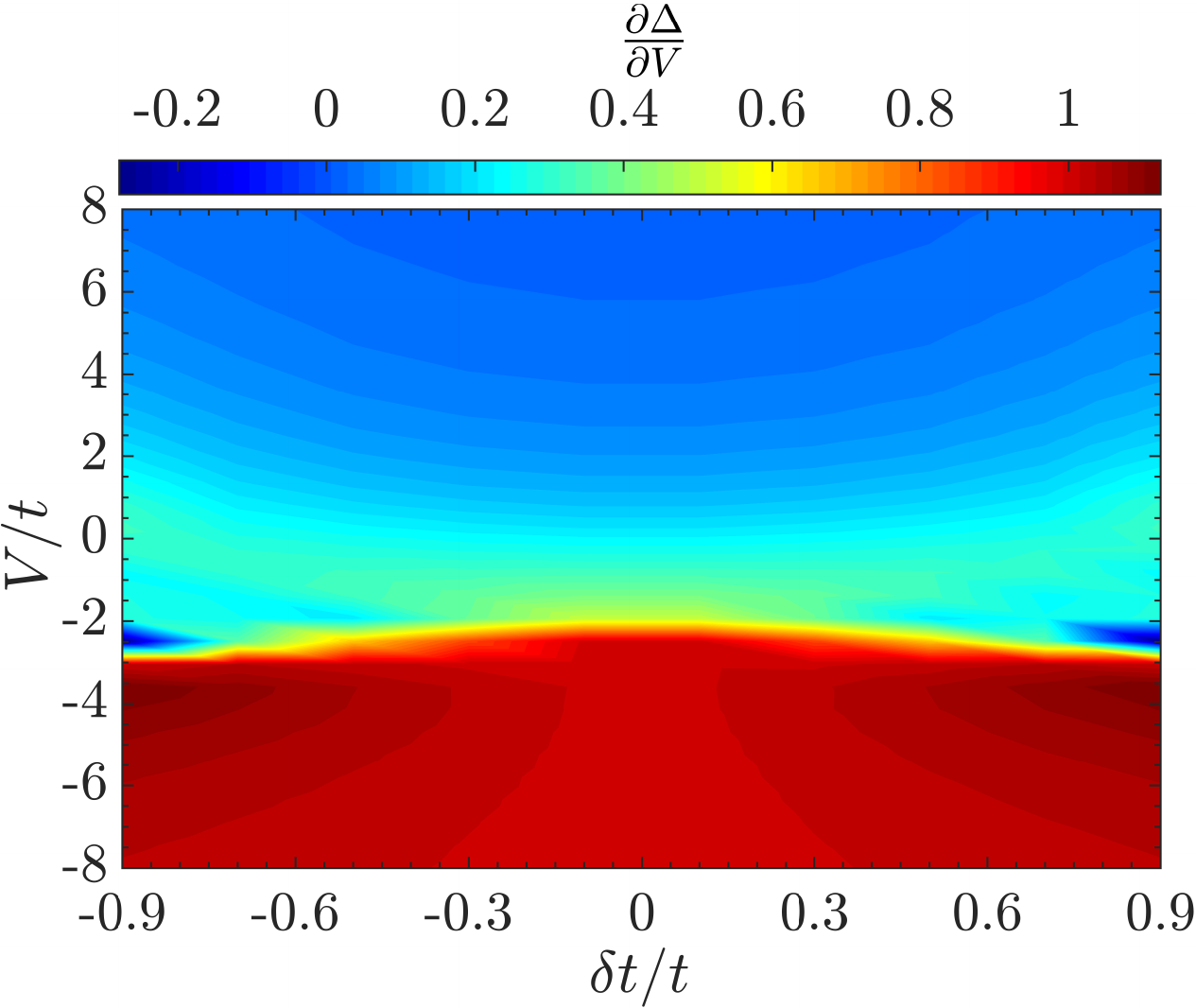}
     \caption{The distribution of the derivative of the energy gap $\Delta$ with respect to $V$ in the parameter space $\left(\delta t,V\right)$.}
     \label{fig4}
   \end{figure}

To more intuitively reveal the sensitivity of the $\Delta$ to $V$, we compute the first derivative of $\Delta$ with respect to $V$ using the finite difference formula:$\frac{\partial\Delta}{\partial V}=\frac{\Delta(V+\delta V)-\Delta(V)}{\delta V}$,where $\delta V=0.5$. The distribution of this derivative across the full parameter space is plotted in Fig.~\ref{fig4}. Notably, a yellow boundary line (with its apex located approximately at $V/t=-2.3$) divides the parameter space into two distinct regions, suggesting a phase transition in the system near this boundary. The downward-opening shape of this boundary line is consistent with the critical values of $V/t$ at which slope discontinuities occur in Fig.~\ref{fig3}(b).
\subsection{\label{subsec:level2}Charge-density structure factor and von Neumann entanglement entropy}

 \begin{figure*}
    \centering
    \includegraphics[width=0.9\textwidth]{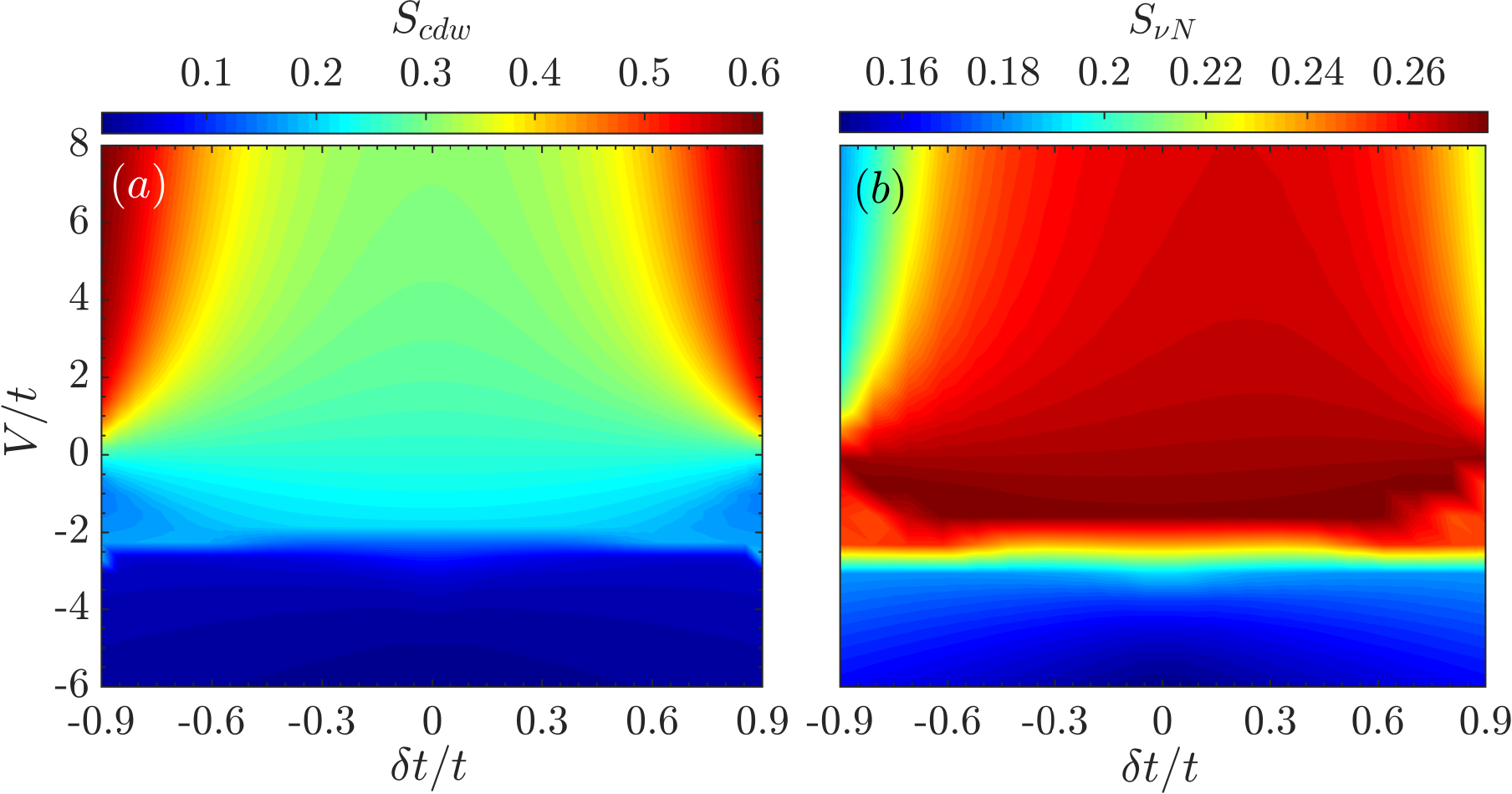}
    \caption{$\left(a\right)$ shows the distribution of the CDW structure factor $S_{cdw}$ in the parameter space $\left(\delta t,V\right)$ ; and $\left(b\right)$ shows the distribution of $S_{\nu N}$.}
    \label{fig5}  
  \end{figure*}  

By comparing Fig.~\ref{fig5}(a) with Fig.~\ref{fig5}(b), one can observe that in certain parameter regions (i.e., in regions with high $S_{cdw}$), a correspondingly high von Neumann entropy $S_{\nu N}$ is present. This is contrary to the expectation in a classical ordered state, where entropy is typically suppressed. This phenomenon indicates that quantum fluctuations in the CDW-ordered phase play a “premelting” role—that is, the fluctuations disrupt the strict long-range order, leading to enhanced quantum entanglement. Furthermore, near $V/t\simeq-2.3$, both $S_{cdw}$ and $S_{\nu N}$ exhibit synchronous sharp transitions, as evidenced by the abrupt color shifts in Fig.~\ref{fig5}(a-b). This indicates a phase transition occurring around $V/t\simeq-2.3$, where the region with $V/t>-2.3$ likely corresponds to the CDW phase, consistent with the yellow line in Fig.~\ref{fig4}.
\subsection{\label{subsec:level3}Destribution of particle}
   \begin{figure*}
      \centering
      \includegraphics[width=0.9\textwidth]{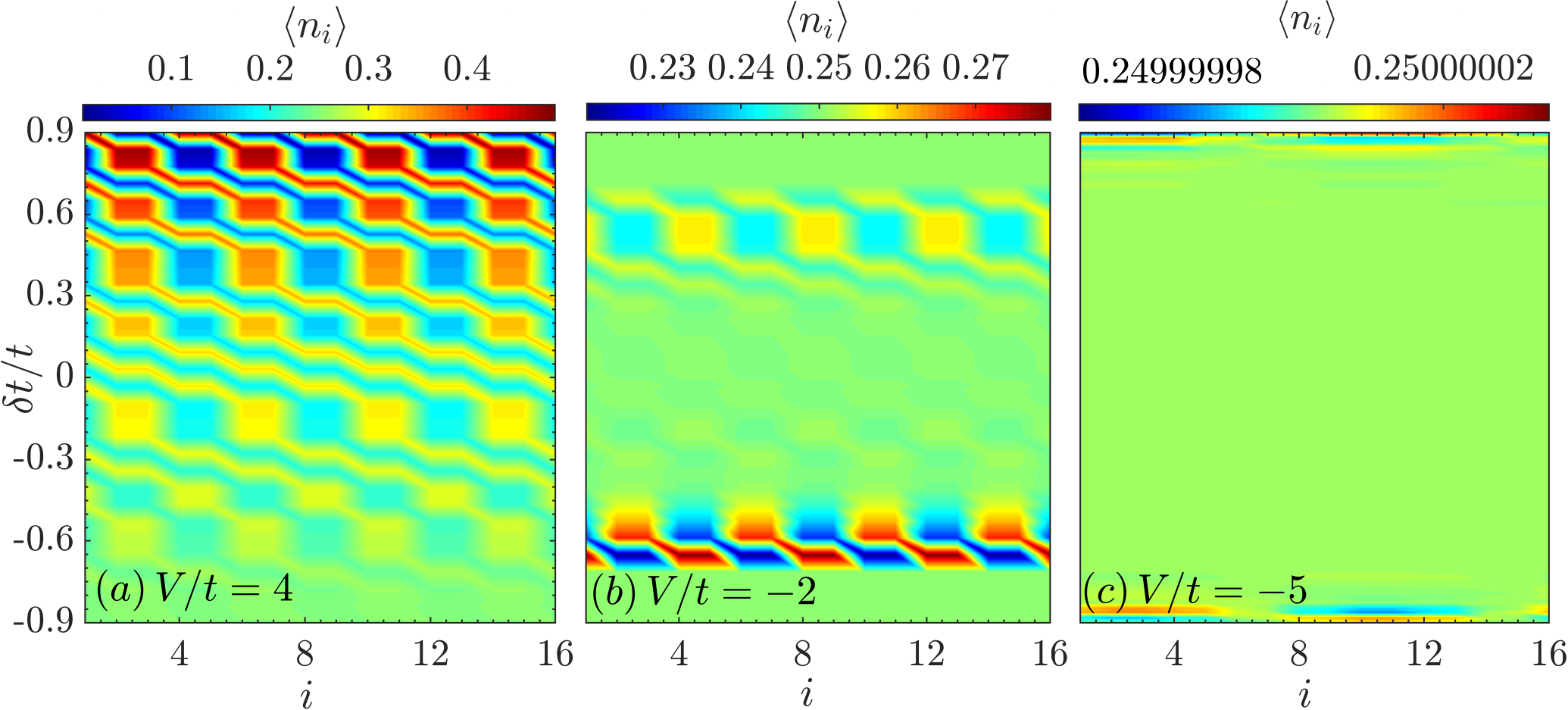}
      \caption{(a),(b) and (c) show, respectively, the distributions of the average occupancy per lattice site in real space under fixed conditions of $V/t=4$, $V/t=-2$ and $V/t=-5$. The subscript $i$ denotes the i-th lattice site.}
      \label{fig6}
   \end{figure*}  
   
By fixing different interaction strengths $V/t$ and calculating the average occupancy distribution of each lattice site over the entire dimerization parameter range $\delta t/t$ (see Fig.~\ref{fig6}), one can observe the following:

In Fig.~\ref{fig6}(a), at $V/t=4$, the average occupancy distribution develops a four-site periodicity, characteristic of a CDW phase. This corresponds to spontaneous breaking of the original single-site translational symmetry, forming a 4-site superlattice structure. Such behavior emerges when the repulsive interaction $V$ dominates over the kinetic energy scale $t$, driving the system into a Wigner-crystalline state with locked commensurate density modulation. 
In Fig.~\ref{fig6}(b), when the interaction is fixed at $V/t=-2$, the system's occupancy exhibits different phase transition behavior. For $\left|\delta t/t\right|<0.71$, the occupancy still shows a modulation with a period of four lattice sites; however, when $\left|\delta t/t\right|>0.71$, the occupancy becomes uniformly distributed. Thus, at $V/t=-2$, the system undergoes a transition from a phase with periodic modulation to a uniform phase protected by topological symmetry.
In Fig.~\ref{fig6}(c), when the interaction is fixed at $V/t=-5$, the occupancy remains uniformly distributed over the entire range of the $\delta t/t$.
  \begin{figure}
      \centering
      \includegraphics[width=0.48\textwidth]{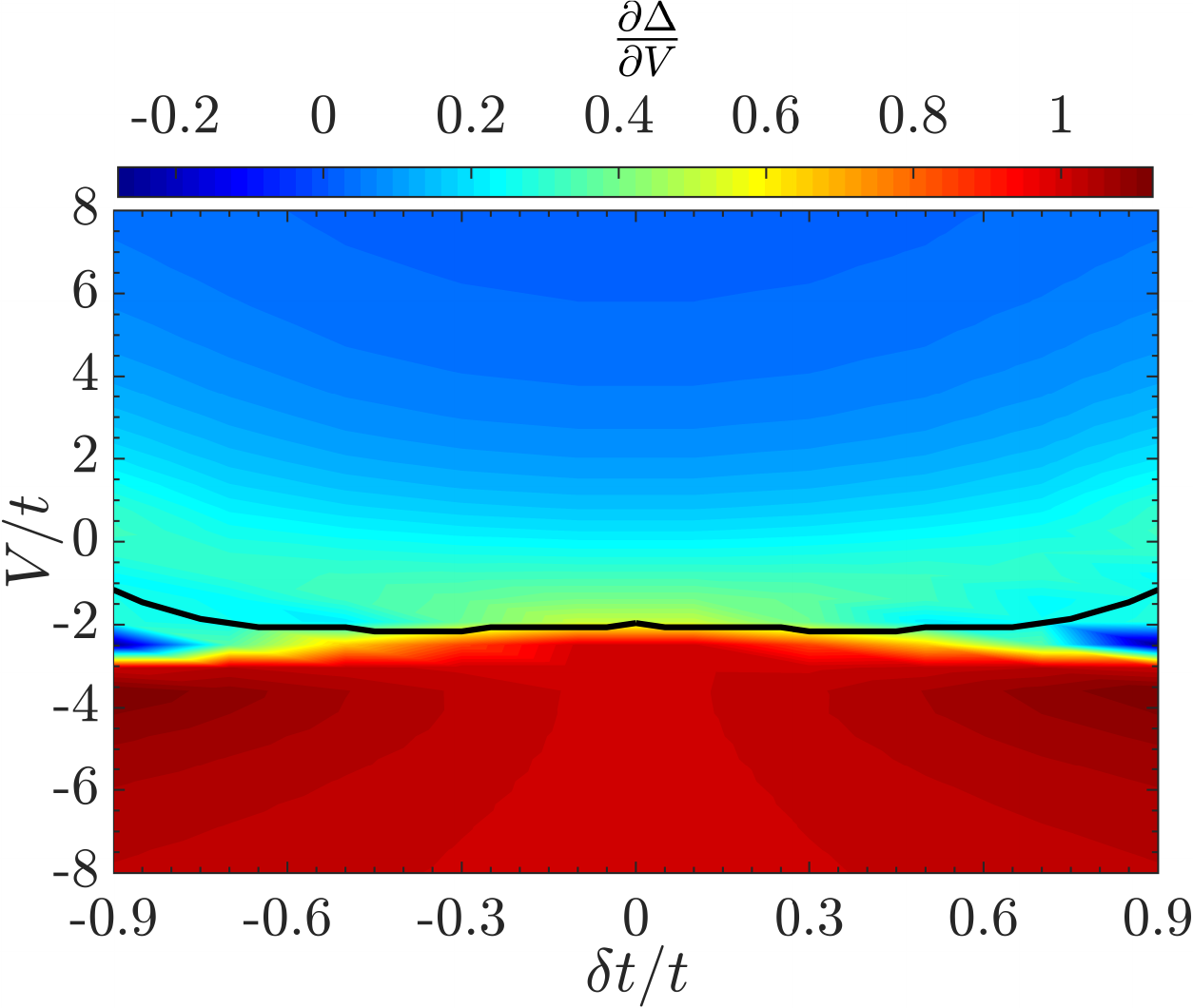}
      \caption{In the parameter space $\left(\delta t,V\right)$, the distribution of the derivative of the $\Delta$ with respect to V is represented by the color scale. Superimposed on this distribution is the critical boundary of the fourfold periodic charge order (black solid line). }
      \label{fig7}
  \end{figure}   
    
By identifying the critical parameters at which changes occur in the average occupancy per lattice site, and plotting these as black solid line (as shown in Fig.~\ref{fig7}). Crossing the line indicates a quantum phase transition from a uniform quarter-filled state to a four-period charge-modulated phase. The essence of this phase transition is a hierarchy of translation symmetry-breaking induced by correlations, closely related to the competitive optimization of the many-body ground state. 
  
Through the complementary verification of the aforementioned physical quantities, we ascertain that the region above the critical line of the average occupancy distribution (depicted as black solid line in Fig.~\ref{fig7}) corresponds to the CDW phase. This observation aligns with the regions of high values for the CDW structure factor $S_{cdw}$ and the von Neumann entropy $S_{\nu N}$ calculated in Sec.~\ref{subsec:level2}. Conversely, in the region below the critical line, the system exhibits a uniform occupancy distribution. Additionally, the area enclosed by the black solid line and the yellow line in Fig.~\ref{fig7} may correspond to a state distinct from the phase region below the yellow line, warranting further analysis to elucidate its specific physical properties.
\subsection{\label{subsec:level4}Competition between CDW and BOW}
 \begin{figure*}
      \centering
      \includegraphics[width=0.9\textwidth]{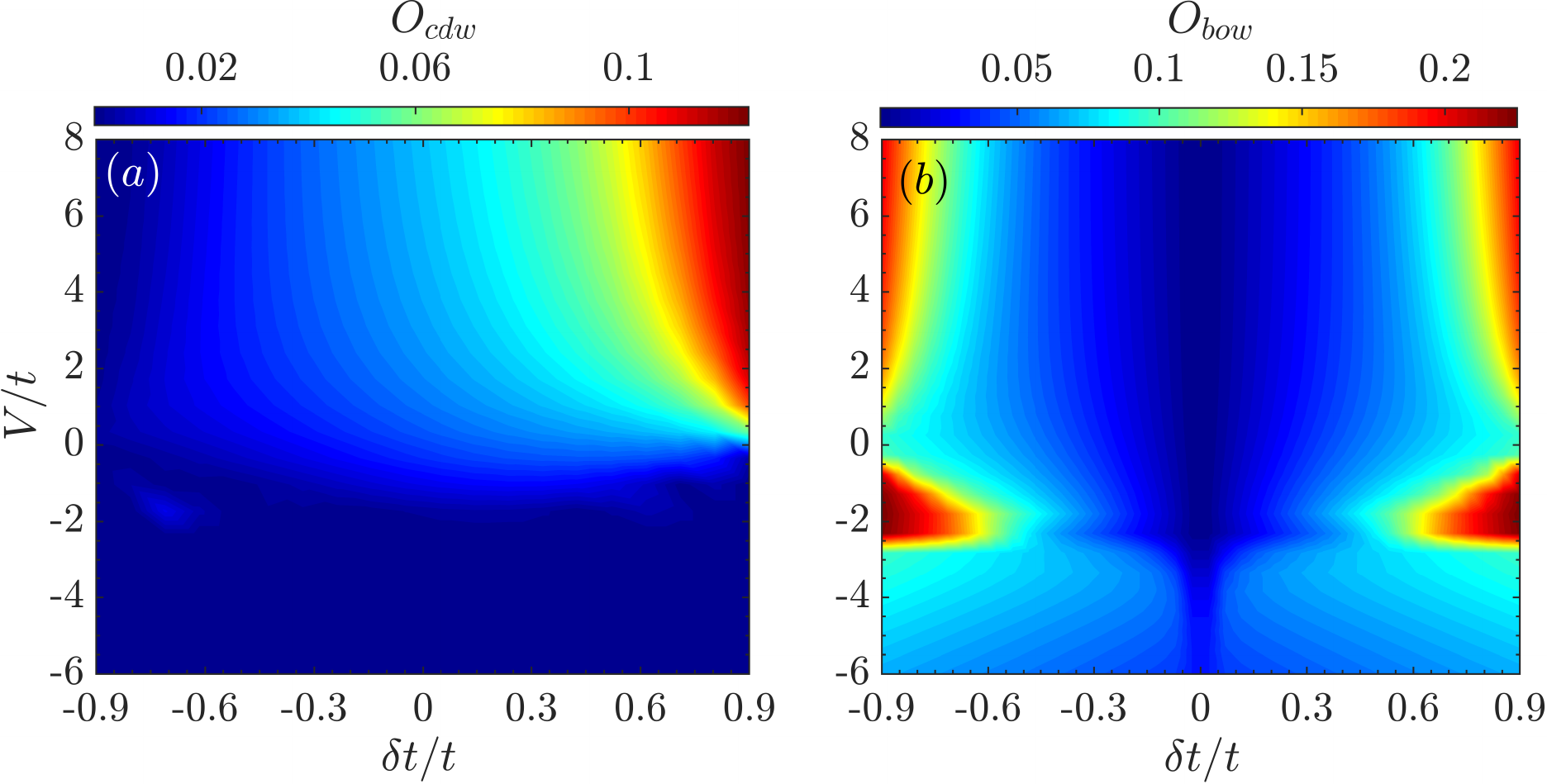}
      \caption{$\left(a\right)$ and $\left(b\right)$ respectively depict the distributions of the CDW amplitude $O_{cdw}$ and the BOW amplitude $O_{bow}$ in the $\left(\delta t,V\right)$ parameter space.}
      \label{fig8}
   \end{figure*}  
   
\begin{figure}
      \centering
      \includegraphics[width=0.48\textwidth]{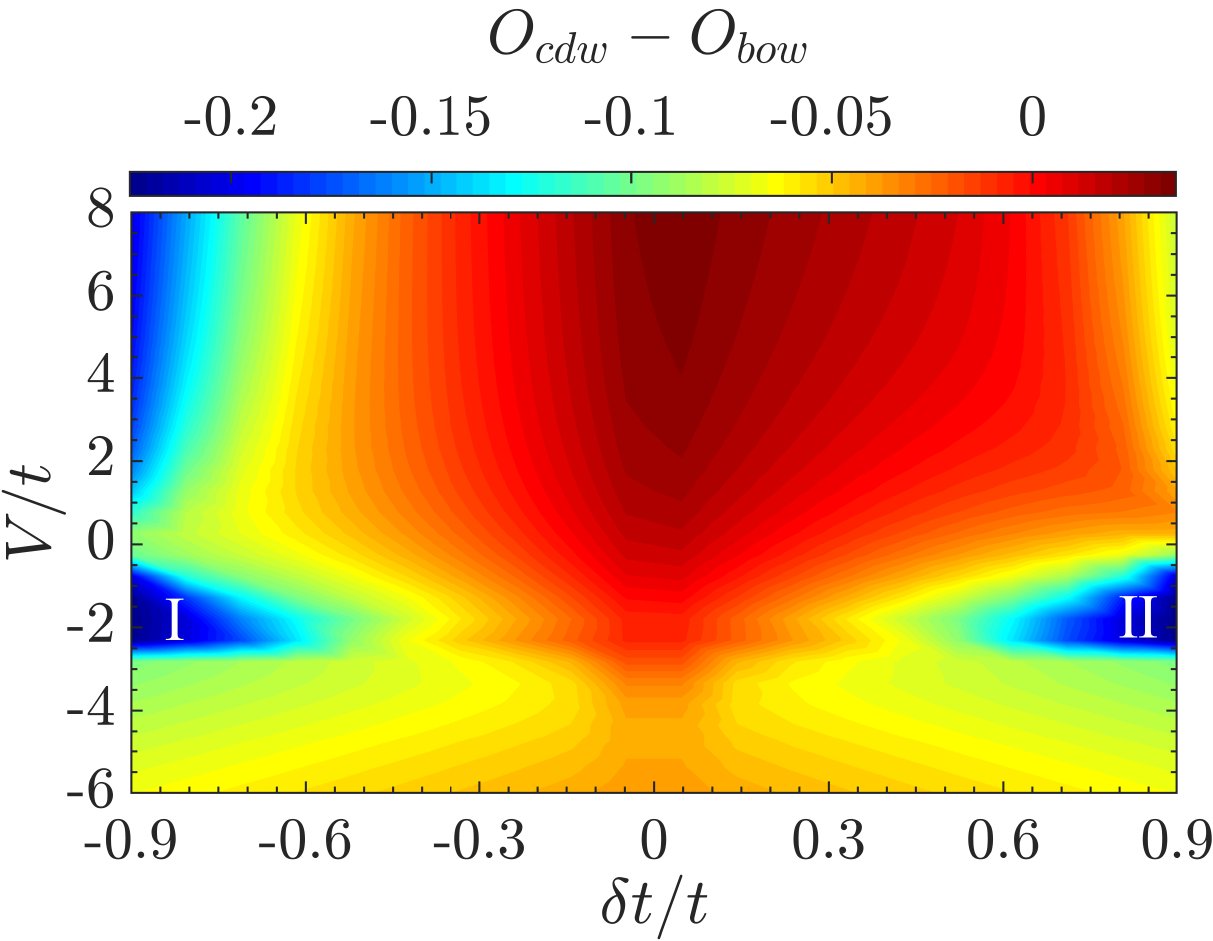}
      \caption{The color map of $\delta = O_{cdw} - O_{bow}$ in the $\left(\delta t,V\right)$ parameter space. }
      \label{fig9}
  \end{figure}

In this work, we investigate the competitive behavior between CDW and BOW through the analysis of order parameters. Fig.~\ref{fig8}(a) and \ref{fig8}(b) respectively present the distribution characteristics  of the CDW and BOW order parameters in the $\left(\delta t,V\right)$ parameter space. Here, the magnitude of the order parameter represents the condensation strength of the corresponding quantum ordered state. To quantitatively determine the dominant regions of these two ordered states, Fig.~\ref{fig9} further constructs the spatial distribution of the difference field $\delta =O_{cdw}-O_{bow}$ between the CDW and BOW order parameters.
  
Analyzing the physical significance of the order parameter difference, a positive $\delta$ indicates that the system is in a CDW-dominated phase, while a negative $\delta$ corresponds to a BOW-dominated phase. Notably, in regions I and II of the Fig.~\ref{fig9}, a significant negative distribution $\left(\delta \ll0\right)$ is observed, indicating that the BOW order parameter amplitude significantly surpasses that of the CDW. According to the competition theory of order parameters in quantum many-body systems, these two regions should be identified as BOW thermodynamic phases. 

Additionally, we observe that the boundaries of regions I and II in Fig.~\ref{fig9} exhibit a high degree of spatial correlation with the areas enclosed by the black solid line and yellow line in Fig.~\ref{fig7}. This phenomenon indicates that the competition between the BOW and CDW phases is not only reflected in the comparison of order parameter magnitudes but also profoundly manifests in the coupling mechanism between the charge spatial order and topological properties of the system's ground state. Furthermore, it can be inferred that the region below the yellow line in Fig.~\ref{fig7} corresponds to the BI phase.
  \begin{figure*}
     \centering
     \includegraphics[width=0.9\textwidth]{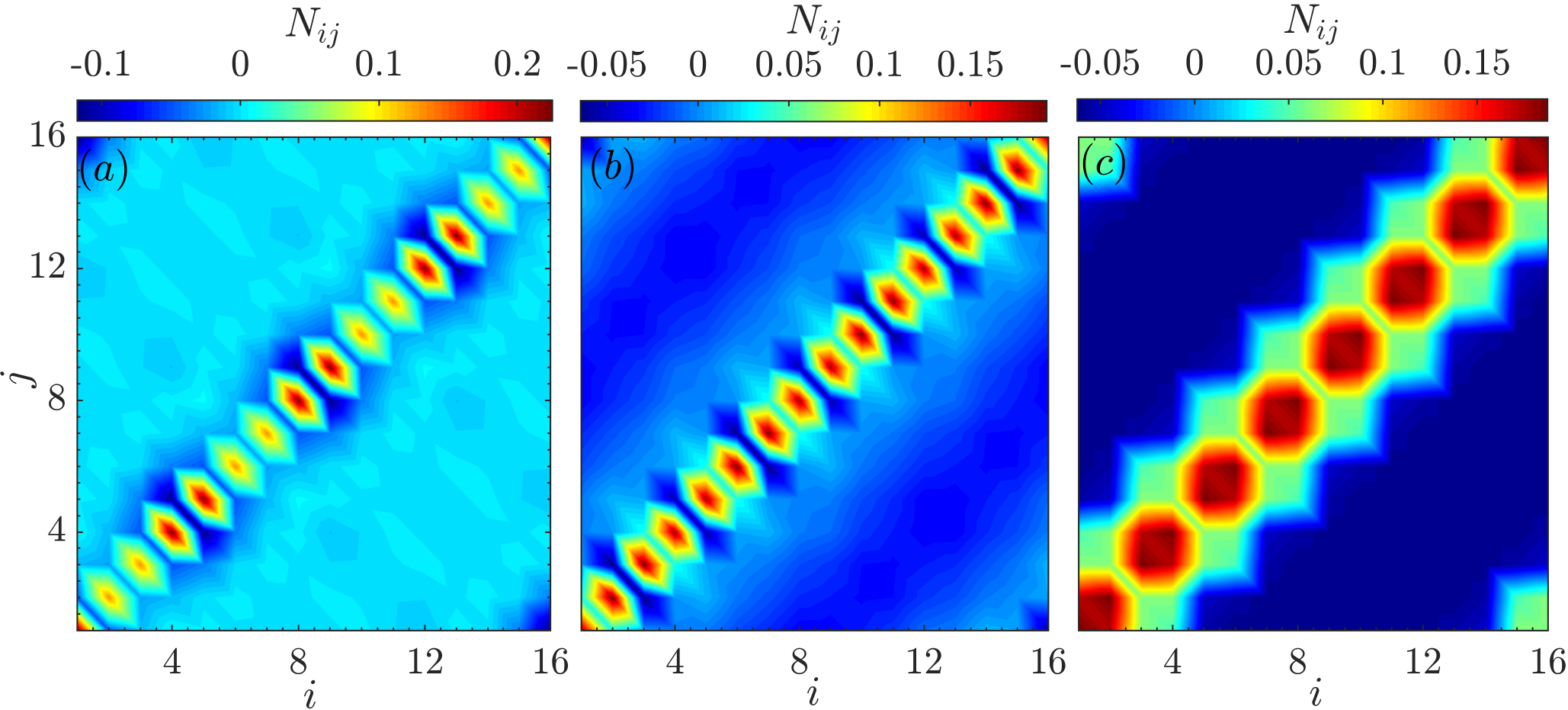}
     \caption{Figures (a), (b), and (c) depict the spatial distribution of the particle-number correlation function $N_{ij}$ in real space, with fixed parameters $\delta t/t=0.3,V/t=4$;$\delta t/t=0.7,V/t=-1.5$and $\delta t/t=0.3,V/t=-5$, respectively. And the subscript $i$ denotes the i-th lattice site. }
     \label{fig10}
  \end{figure*}  

To better illustrate the characteristics among different phases, we also calculate the real-space particle number correlation function (Eq.~\ref{eq8}), with results shown in Fig.~\ref{fig10}. Panels (a), (b), and (c) of Fig.~\ref{fig10} correspond to parameter choices that respectively represent the CDW, BOW, and BI phases. From the plots, we observe that panel (a) exhibits a strong periodic charge modulation with a period of 4, indicating a long-range ordered CDW phase. Panel (b) shows alternating positive and negative correlations without strict periodicity, suggesting a BOW phase with bond modulation. In panel (c), the correlations weaken and become more uniform, corresponding to the featureless BI phase.
\subsection{\label{subsec:level5}Phase boundary of the BOW phase}
\begin{figure*}
      \centering
      \includegraphics[width=0.9\textwidth]{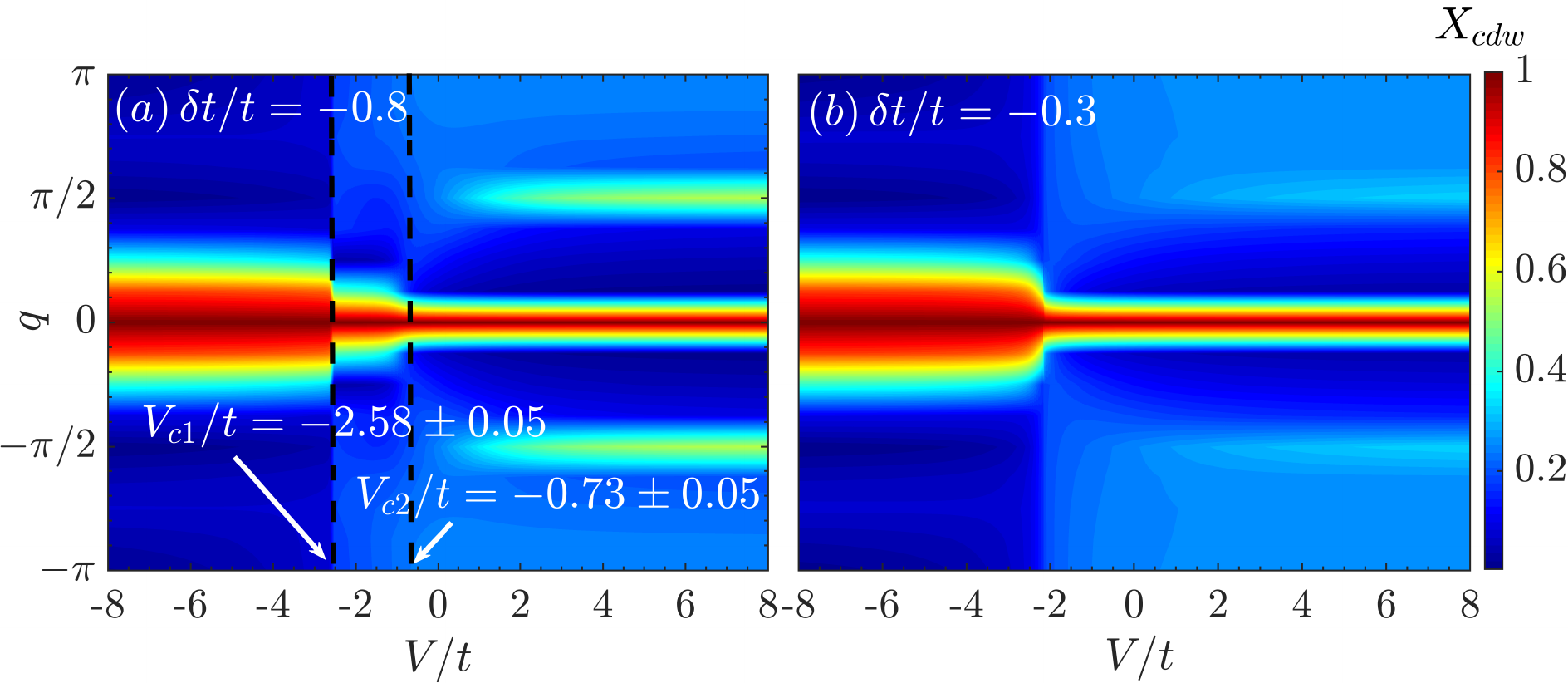}
      \caption{$\left(a\right)$ and $\left(b\right)$ display the color maps of the CDW correlation function $\chi_{cdw}(q)$ in momentum space, showing its distribution within the $\left(V,q\right)$ parameter space for different fixed values of $\delta t /t$. }
      \label{fig11}
   \end{figure*}  
   
This study utilizes momentum-space correlation function spectroscopy to elucidate the characteristic responses of BI, BOW, and CDW phases under the influence of electron correlations. Fig.~\ref{fig11} presents the distribution features of the CDW correlation function $\chi_{cdw}(q)$ within the parameter space $\left(\delta t,V\right)$. The evolution of singular structures in momentum space directly reflects the symmetry-breaking patterns of different quantum phases.

When $V$ is less than a critical value $V_{c1}$, the CDW correlation function, $\chi_{cdw}(q)$, exhibits a pronounced peak exclusively at $q=0$ in momentum space (see Fig.~\ref{fig11}(a)). This behavior arises from the uniform ground state with unbroken translational symmetry: the enhanced correlation at $q=0$ corresponds to the suppression of overall charge fluctuations in the system. This is consistent with the characteristics of BI phase, where electron localization leads to the absence of long-range correlations in momentum space.

In the region $V_{c1} < V < V_{c2}$, since the primary feature of the BOW phase is the staggered modulation of hopping amplitudes rather than direct charge density fluctuations, $\chi_{cdw}(q)$ still exhibit a significant peak at $q=0$, similar to the BI phase.However, the behavior of $\chi_{cdw}(q)$ in these two phases still differs, allowing them to be distinguished from each other.

When the interaction strength $V$ exceeds a critical value $V_{c2}$, $\chi_{cdw}(q)$ exhibits pronounced peaks at $q = 0$, $\pm\frac{\pi}{2}$ in momentum space. This multi-peak structure reveals the essence of the fourfold periodic modulation in the CDW phase, characterized by a wave vector $Q = \frac{\pi}{2}$. This behavior can be partially attributed to the nesting effect at quarter filling, where the interaction-driven Fermi surface instability favors charge ordering at $Q = \frac{\pi}{2}$. However, in the strong interaction regime, charge localization effects also play a crucial role in stabilizing the CDW phase. 

Additionally, by comparing panels (a) and (b) of Fig.~\ref{fig11}, it is clear that no BOW phase exists when $\delta t/t=-0.3$.
   \begin{figure*}
      \centering
      \includegraphics[width=0.9\textwidth]{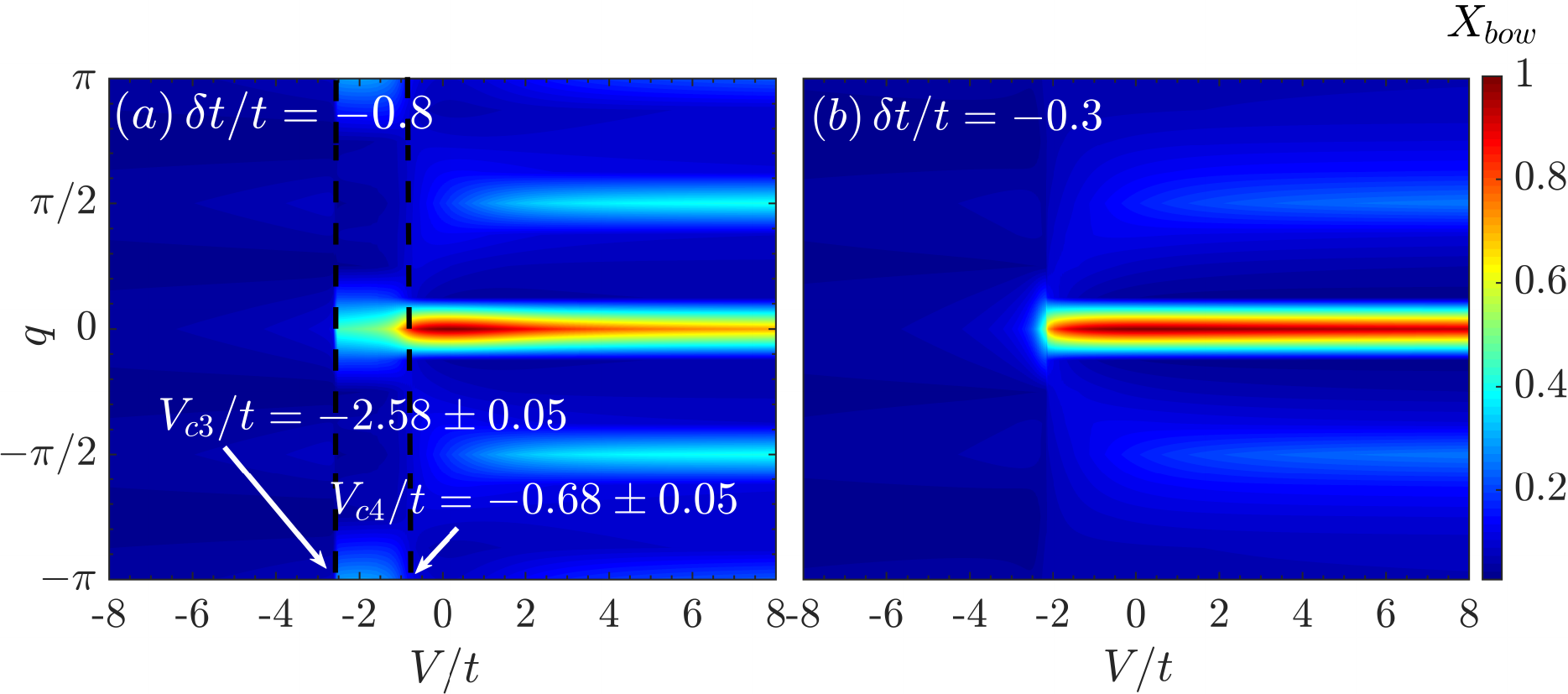}
      \caption{$\left(a\right)$ and $\left(b\right)$ display the color maps of the BOW correlation function $\chi_{bow}(q)$ in momentum space, showing its distribution within the $\left(V,q\right)$ parameter space for different fixed values of $\delta t /t$.}
      \label{fig12}
   \end{figure*}  

Fig.~\ref{fig12} illustrates the evolution of the BOW correlation function, $\chi_{bow}(q)$, across different regions of the parameter space. Observing Fig.~\ref{fig12}(a), in the BI phase characterized by $V < V_{c3}$, $\chi_{bow}(q)$ exhibits negligible oscillations, indicating the absence of long-range bond order. This coupled with the single peak of $\chi_{cdw}$ at $q=0$, corroborates the presence of a uniform ground state with unbroken translational symmetry.

In the BOW phase ($V_{c3} < V < V_{c4}$), the correlation function $\chi_{bow}(q)$ shows significant peaks at $q=0$ and $q=\pm\pi$. The peaks at $q=\pm\pi$ indicate the establishment of long-range dimerized bond order, corresponding to the breaking of lattice translational symmetry to a doubled unit cell. This phenomenon is analogous to the topologically nontrivial state in the SSH model, though here the bond order is driven by electronic correlations rather than static lattice distortions. This feature complements the dual peaks of $\chi_{cdw}$ only at $q=0$  observed in the BOW phase. 

In the CDW phase ($V > V_{c4}$), $\chi_{bow}$ exhibits pronounced peaks at $q = \pm \frac{\pi}{2}$, which coincide with the enhanced peaks in the charge correlation function $\chi_{cdw}$ at the same wave vectors. This observation indicates a strong coupling between bond order and charge order in the CDW phase. Specifically, the fourfold periodic modulation of the charge density (with wave vector $Q = \frac{\pi}{2}$) induces corresponding modulations in the bond lengths through electron-lattice coupling. Although theoretical models predict that such nonlinear coupling could generate higher harmonics (e.g., $2Q = \pi$ and $4Q = 2\pi \equiv 0$), in practice the most prominent resonant response associated with the CDW ordering is observed at $q = \pm \frac{\pi}{2}$. The $q = 0$ component reflects the overall charge fluctuations, while the expected response at $q = \pi$ is either too weak or obscured by the dominant $q = \pm \frac{\pi}{2}$ signal. 

By comparing Fig.~\ref{fig12}(a) and  \ref{fig12}(b), we observe that when $\delta t = -0.3$, the system does not exhibit a BOW phase.

We simultaneously compare and discuss $\chi_{cdw}$ and $\chi_{bow}$. $\chi_{cdw}$ is sensitive to the charge degrees of freedom, its multi-peak structure—such as the $\pm\frac{\pi}{2}$ peaks observed in the CDW phase—directly reflects the modulation period resulting from broken translational symmetry. Conversely, the peak positions of $\chi_{bow}$ reveal the evolution of bond order. In the BOW phase, the intensity of the $\pm\pi$ peaks in $X_{bow}$ is significantly higher than in $\chi_{cdw}$, indicating that bond order is the dominant order parameter. In the CDW phase, both $\chi_{bow}$ and $\chi_{cdw}$ exhibit $\pm\frac{\pi}{2}$ peaks; however, $\chi_{cdw}$ dominates, corroborating the notion that charge order drives bond order.
   \begin{figure}
      \centering
      \includegraphics[width=0.48\textwidth,height=0.3\textheight]{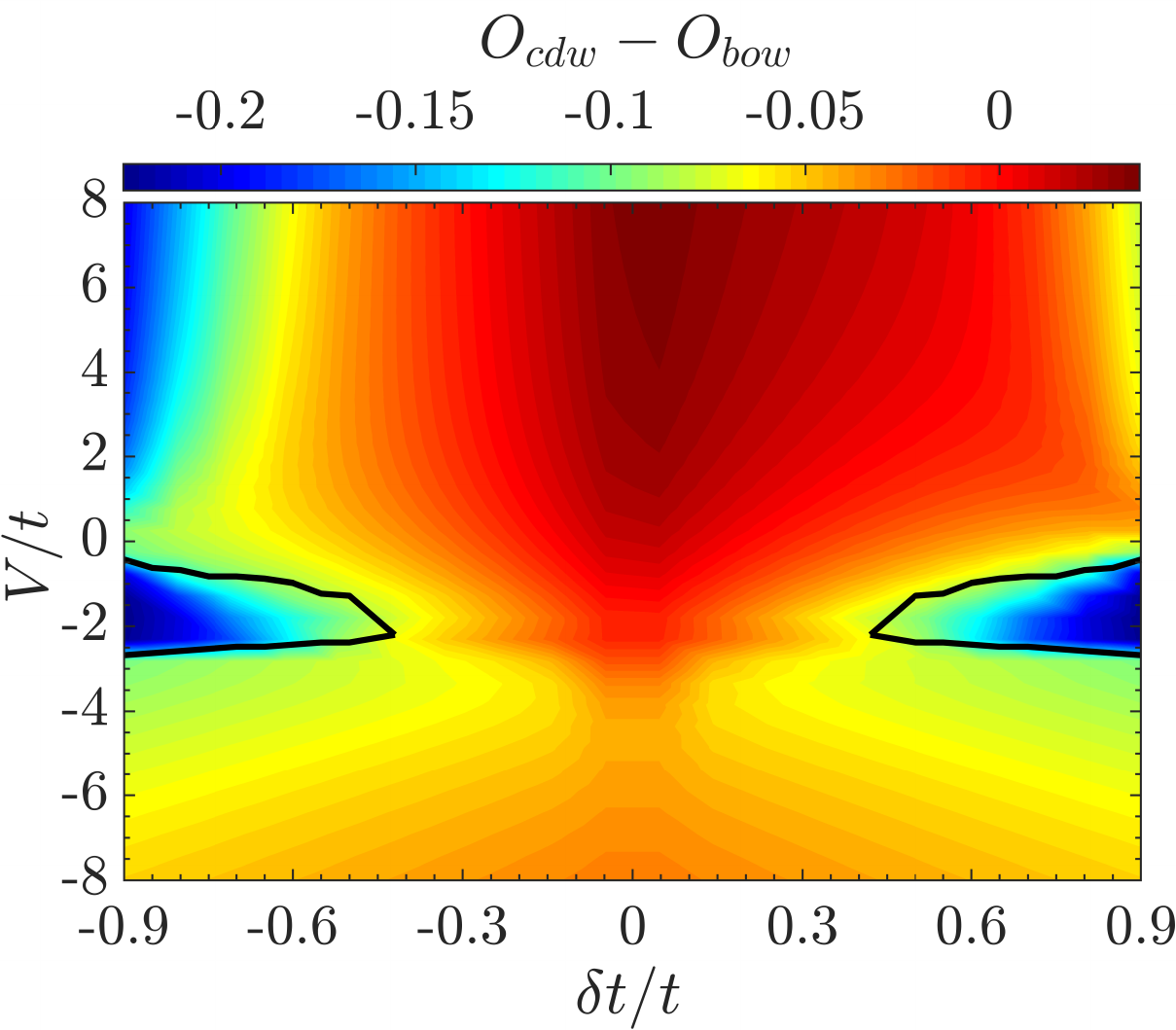}
      \caption{The figure displays the phase boundary of the BOW phase, represented by a solid black line, superimposed on a color gradient map illustrating the amplitude of the parameter $\delta=O_{cdw}-O_{bow}$. }
      \label{fig13}
   \end{figure}  
   
By analyzing the $\chi_{bow}(q)$, we can determine the critical interaction strength $V_c$ and the corresponding hopping parameter $\delta t$ that signify the presence of the BOW phase. Overlaying the BOW phase boundaries onto the distribution map of the amplitude difference $\delta = O_{cdw} - O_{bow}$ (as shown in Fig.~\ref{fig13}), we observe that within the BOW phase boundaries, the BOW amplitude is significantly greater than that of the CDW phase. This indicates that within this region, the order parameter of the BOW state is dominant, reflecting the system's preference for BOW formation during spontaneous symmetry breaking. These observations, corroborated by various physical quantities, collectively and comprehensively affirm the intrinsic characteristics of the BOW phase.

  \begin{figure}
      \centering
      \includegraphics[width=0.48\textwidth,height=0.3\textheight]{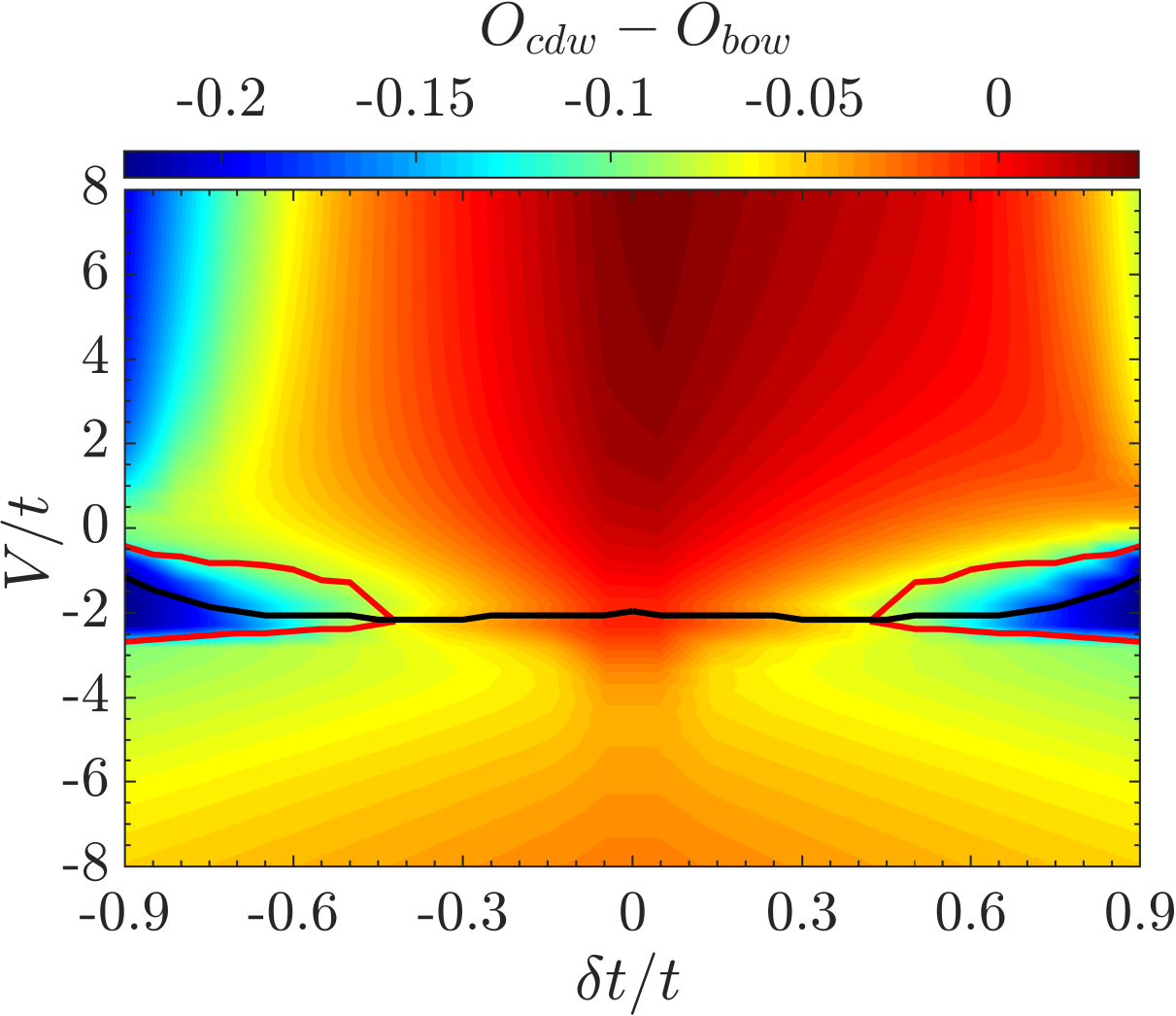}
      \caption{The figure displays the phase boundary of the BOW phase (solid red line), and the critical boundary of the fourfold periodic charge order (solid black line), all superimposed on a color gradient map illustrating the amplitude of the parameter $\delta =O_{cdw}-O_{bow}$. }
      \label{fig14}
   \end{figure}  
   
Although the upper boundary of the BOW phase, determined via the function $\chi_{bow}$, exhibits strong spatial consistency with the critical line obtained from the average site occupancy distribution, noticeable numerical discrepancies are still present in Fig.~\ref{fig14}. These deviations may arise from several sources. First, $\chi_{bow}$ primarily captures long-range bond-order correlations, whereas the average occupancy reflects the local distribution of particles across lattice sites. The distinct physical quantities they probe result in differing sensitivities to variations in system parameters, potentially leading to mismatches in the identified critical points. Second, numerical simulations are typically performed on finite-sized systems, where finite-size effects can differently impact correlation functions and local observables, leading to shifts in the extracted critical points. Furthermore, rounding errors and the limitations of numerical precision inherent in computational algorithms may introduce additional deviations. Therefore, although the critical lines derived from  $\chi_{bow}$ and the average occupancy distribution exhibit consistent trends, numerical differences remain due to the aforementioned factors.
\section{\label{sec:level4}Compared to the mean-field approximation results} 
Building upon the previous ED results, we have shown that the one-dimensional spinless SSH model at quarter filling exhibits distinct phases, including the BOW, CDW, and BI phases. To validate these findings, we apply a mean-field approximation to the original Hamiltonian. Considering the possibility of both BOW and CDW phases, we adopt two distinct types of mean-field decouplings.

We begin with the BOW-type mean-field approximation, and rewrite the interaction term as:
    \begin{equation}\label{eq14}
    \begin{split}
    \hat{n}_i\hat{n}_j &= \hat{c}_i^\dagger\hat{c}_i\hat{c}_j^\dagger\hat{c}_j = -\hat{c}_i^\dagger\hat{c}_j^\dagger\hat{c}_i\hat{c}_j \\
    &\simeq -\langle\hat{c}_i^\dagger\hat{c}_j\rangle\hat{c}_j^\dagger\hat{c}_i
    - \langle\hat{c}_j^\dagger\hat{c}_i\rangle\hat{c}_i^\dagger\hat{c}_j
    + \langle\hat{c}_i^\dagger\hat{c}_j\rangle\langle\hat{c}_j^\dagger\hat{c}_i\rangle,
    \end{split}
    \end{equation}
by defining $\chi_{ij}=\langle\hat{c}_j^\dagger\hat{c}_i\rangle=\chi+(-1)^i\delta\chi$, the interaction term becomes:
   \begin{equation}\label{eq15}
       \hat{n}_i\hat{n}_j\simeq-\chi_{ij}^*\hat{c}_j^\dagger\hat{c}_i-\chi_{ij}\hat{c}_i^\dagger\hat{c}_j+|\chi_{ij}|^2.
    \end{equation}
Substituting Eq.~\ref{eq15} into the Eq.~\ref{eq1} yields the BOW-type mean-field Hamiltonian:
   \begin{equation}\label{eq16}
       \hat{H}_{MF}=\sum_{\langle i,j\rangle}(-t+(-1)^i\delta t-V\chi_{ij})(\hat{c}_j^\dagger\hat{c}_i+h.c.)+V\sum_{\langle i,j\rangle}|\chi_{ij}|^2.
    \end{equation}
The presence of a BOW phase is determined by the condition $\delta\chi \neq 0$, while $\delta\chi = 0$ indicates the absence of BOW order in the system.
   
Similarly, we apply a CDW-type mean-field approximation. The interaction term is approximated as
   \begin{equation}\label{eq17}
       \hat{n}_i\hat{n}_j\simeq\hat{n}_i\langle n_j\rangle+\langle n_i\rangle\hat{n}_j-\langle n_i\rangle\langle n_j\rangle,
   \end{equation}
where we define the average site occupation as $\langle n_i\rangle=n+e^{iqR_i}\delta n, q=\frac{\pi}{2}$. And we can obtain
   \begin{equation}\label{eq18}
   \begin{split}
    \hat{n}_i\hat{n}_j 
    &\simeq \hat{n}_i(n+e^{i\frac{\pi}{2}R_j}\delta n) + (n+e^{i\frac{\pi}{2}R_i}\delta n)\hat{n}_j \\
    &- (n +e^{i\frac{\pi}{2}R_i} \delta n)(n +e^{i\frac{\pi}{2}R_j} \delta n) \\
    &= n(\hat{n}_i+\hat{n}_j)+\delta n(\hat{n}_ie^{i\frac{\pi}{2}R_j}+\hat{n}_je^{i\frac{\pi}{2}R_i})\\
    &-[n^2+n\delta n(e^{i\frac{\pi}{2}R_i}+e^{i\frac{\pi}{2}R_j})+(\delta n)^2e^{i\frac{\pi}{2}(R_i+R_j)}].
   \end{split}
   \end{equation}
However, we find that summing over all nearest-neighbor pairs yields
  \begin{equation}\label{eq19}
    V\sum_{\langle i,j\rangle}\hat{n}_i\hat{n}_j =Vzn\sum_{i}\hat{n}_i-\frac{VzL}{2}n^2,
  \end{equation}
where $z = 2$ is the coordination number in the one-dimensional SSH model. Notably, the final expression contains no terms involving the order parameter $\delta n$, implying that $\delta n$ cannot serve as an indicator of CDW order within this mean-field framework. This absence does not necessarily indicate that the true ground state lacks CDW character; rather, it reflects a fundamental limitation of mean-field theory in capturing strong quantum fluctuations, especially in one-dimensional systems where spontaneous symmetry breaking is significantly suppressed.

To overcome this issue and probe the possibility of CDW formation within the MF framework, we introduced a next-nearest-neighbor (NNN) interaction $V_1\sum_{\langle\langle i,j\rangle\rangle}\hat{n}_i\hat{n}_j$, and apply a mean-field approximation. Consequently, the resulting mean-field Hamiltonian under the CDW-type modulation ansatz takes the form
  \begin{equation}\label{20}
   \begin{split}
   \hat{H}_{MF} = &\sum_{\langle i,j\rangle} (-t + (-1)^i \delta t)(\hat{c}_j^\dagger \hat{c}_i + h.c.) \\
    &+ (Vzn+V_1z(n - i\delta n))\sum_{i_1 \in \{i \mid i \bmod 4 = 1\}} \hat{n}_{i_1}\\
    &+ (Vzn+V_1z(n +\delta n))\sum_{i_2 \in \{i \mid i \bmod 4 = 2\}} \hat{n}_{i_2} \\
    &+ (Vzn+V_1z(n +i\delta n))\sum_{i_3 \in \{i \mid i \bmod 4 = 3\}} \hat{n}_{i_3}\\
    &+ (Vzn+V_1z(n -\delta n))\sum_{i_4 \in \{i \mid i \bmod 4 = 0\}} \hat{n}_{i_4} \\
    &-\frac{VzL}{2}n^2- \frac{V_1zL}{2}n^2.
    \end{split}
  \end{equation}
The inclusion of NNN interactions enables the stabilization of a finite $\delta n$, signaling the emergence of a CDW phase. This result aligns with the ED findings and confirms that the system tends toward CDW order. Importantly, the NNN interaction is not introduced to alter the physical model, but rather serves as a technical means for the MF approach to capture the underlying CDW tendency already present in the system.

Similar to the BOW case, a nonzero value of the order parameter $\delta n$ indicates the presence of CDW order, while $\delta n = 0$ corresponds to a uniform phase without charge modulation.

\begin{figure*}
      \centering
      \includegraphics[width=0.9\textwidth]{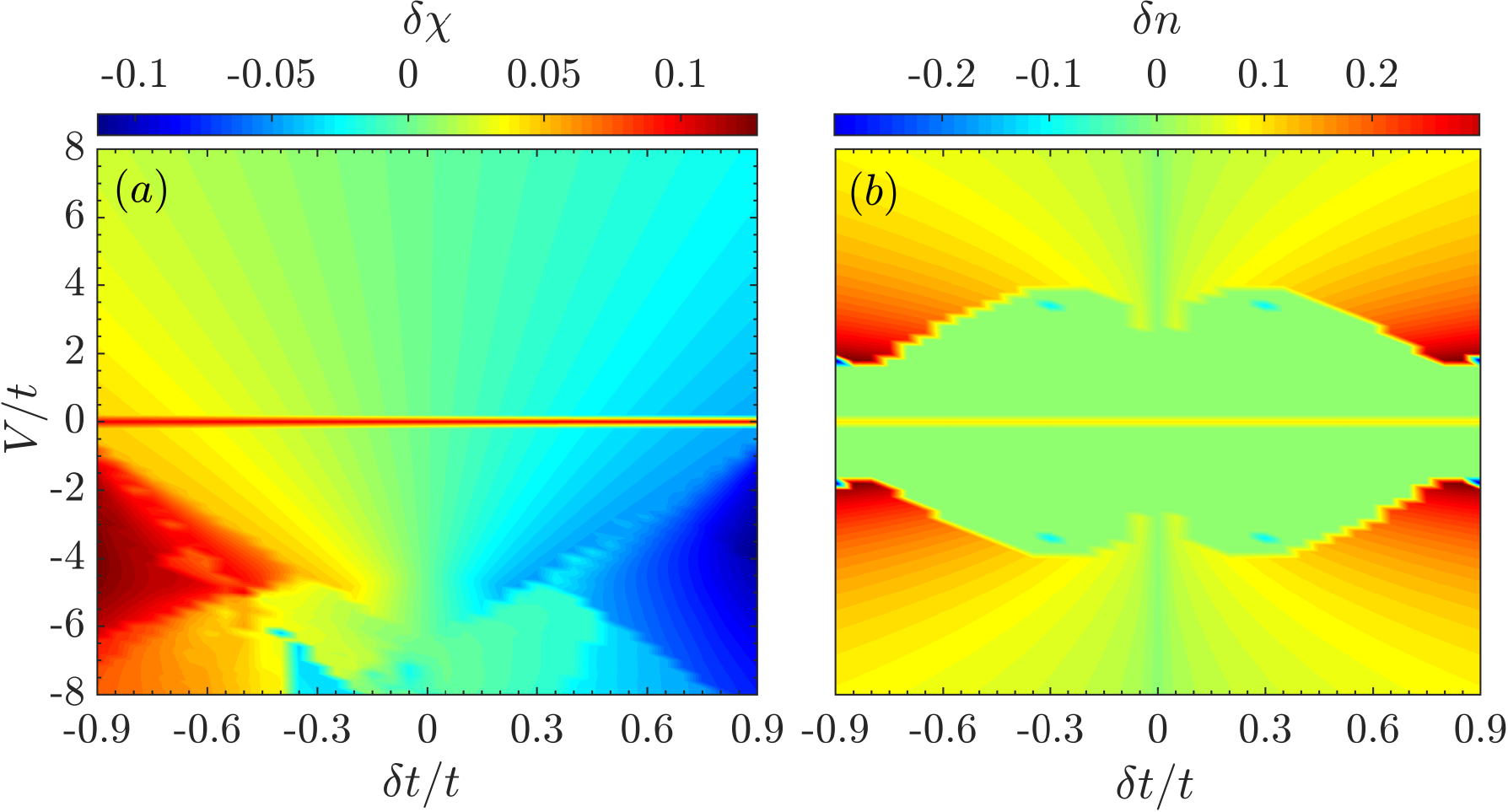}
      \caption{(a) presents the color map of $\delta\chi$ in the parameter space $\left(\delta t,V\right)$ under the BOW-type mean-field approximation, while (b) shows the color map of $\delta n$ in the parameter space $\left(\delta t,V\right)$ under the CDW-type mean-field approximation with the next-nearest-neighbor interaction fixed at $V_1 = V$.}
      \label{fig15}
   \end{figure*}

Based on the above derivations, we obtain two forms of mean-field Hamiltonians corresponding to the BOW-type and CDW-type approximations. We then employ variational wavefunctions to determine the values of the order parameters $\delta\chi$ and $\delta n$ that minimize the ground-state energy under each mean-field approximation. The resulting values are mapped onto the parameter space $(\delta t, V)$, as shown in Fig.~\ref{fig15}. Fig.~\ref{fig15} (a) and (b) show the BOW and CDW phase regions, identified by nonzero $\delta\chi$ and $\delta n$, respectively. It is worth noting that the dependence of the ground-state energy on $\delta n$ appears only through even powers of the NNN interaction strength $V_1$. As a result, the energy landscape and corresponding order parameters exhibit mirror symmetry with respect to the line $V=0$ in panel (b). In this study, we restrict our attention to the $V>0$ region because the formation of a CDW phase typically requires a repulsive interaction between NNN sites. Therefore, a CDW phase is not expected to occur in the $V < 0$ regime.
   
While the mean-field approximation successfully captures the qualitative features of the phase diagram, quantitative discrepancies remain when compared to the ED results. These deviations primarily arise from the intrinsic limitations of the mean-field approach, which neglects quantum fluctuations and many-body correlations beyond the averaged field description. Such effects are especially pronounced in low-dimensional systems like the one considered here, where quantum fluctuations can significantly influence the ground-state properties and lead to noticeable shifts in the location and extent of phase boundaries.
\section{\label{sec:level5}Conclusion}
Through ED and MFA methods, this paper investigates the ground-state phase diagram of a spinless one-dimensional Su-Schrieffer-Heeger model with 16 lattice sites under quarter-filling conditions, incorporating nearest-neighbor interactions. The ED results reveal that the system exhibits a topologically trivial BI phase for strong attractive interactions, with its upper boundary forming a downward-opening curve peaking at $V/t\simeq-2.3$ and extending to $V/t\simeq-2.6$,  as further evidenced by the uniform distribution of the average site occupation. In the range of $-2.6 \leq V/t \leq -0.5$ and $|\delta t/t| > 0.45$, the system transitions into a BOW phase characterized by spontaneous bond-order symmetry breaking. This phase exhibits quasi-long-range modulation of bond strengths, where its stabilization arises from the synergistic interplay between electronic correlations and the intrinsic bond order of the SSH model. In other parameter regimes, the CDW phase dominates, with its quadruple superlattice periodic modulation pattern clearly characterized by Bragg peaks in momentum-space correlation functions. 

Although the MFA inherently involves approximations in treating many-body systems and is less accurate than ED in a quantitative sense, it qualitatively reproduces the phase boundaries revealed by ED through the analysis of bond-order ($\delta \chi$) and charge-density ($\delta n$) modulations. In particular, it confirms the presence and stability of the BOW and CDW phases. The consistency between ED and MFA results further highlights the intrinsic nature of these ordered phases and their robustness against different theoretical approaches.

By synthesizing insights from both ED and MFA, this work constructs a comprehensive phase diagram governed by the competition between distinct order parameters. These findings offer a coherent theoretical framework for understanding the interplay between topology, electronic correlations, and spontaneous symmetry breaking in one-dimensional quantum systems. They further provide valuable guidance for future investigations into topological phases and quantum critical phenomena in low-dimensional correlated materials.
\begin{acknowledgments}
This work was supported by the National Natural Science Foundation of China (Grant No.~12247101), the Fundamental Research Funds for the Central Universities (Grant No.~lzujbky-2024-jdzx06), the Natural Science Foundation of Gansu Province (Grant Nos.~22JR5RA389 and 25JRRA799), and the “111 Center” under Grant No.~B20063.
\end{acknowledgments}

\bibliographystyle{apsrev4-2} 
\bibliography{refs}
\end{document}